\documentclass[usenatbib]{mn2e}
\usepackage[dvips]{graphicx}
\usepackage{colortbl}
\usepackage{amssymb,txfonts}
\usepackage[dvipdfm,colorlinks=true,linkcolor=blue,urlcolor=blue,citecolor=blue]{hyperref}

\newcommand{\apj}{{ApJ}}

\newcommand{\mnras}{{MNRAS}}

\newcommand{\araa}{{ARA\&A}}

\newcommand{\msun}{{\rm M}_{\sun}}

\def\ee{\end{equation}}
\def\be{\begin{equation}}
\newcommand{\ledd}{L_{{\rm Edd}}}
\newcommand{\mdot}{\dot M}

\newcommand{\ergs}{{\rm \,erg\,s^{-1}}}

\title[jet-dominated quiescent states]{jet-dominated quiescent states in black hole X-ray binaries: the case of V404 Cyg}

\author[F. G. Xie, Q. X. Yang \& R. Ma]{Fu-Guo Xie,$^{1}$
Qi-Xiang Yang,$^{1,2}$ Renyi Ma$^{3}$\\
$^1$Key Laboratory for Research in Galaxies and Cosmology, Shanghai Astronomical Observatory, Chinese Academy of Sciences, \\
80 Nandan Road, Shanghai 200030, China; fgxie@shao.ac.cn\\
$^2$University of Chinese Academy of Sciences, 19A Yuquan Road, Beijing 100049, China\\
$^3$Department of Astronomy and Institute of Theoretical Physics and Astrophysics, Xiamen University, Xiamen 361005, Fujian Province, China
}

\date{}
\pagerange{\pageref{firstpage}--\pageref{lastpage}}

\begin{document}

\maketitle

\label{firstpage}

\begin{abstract}
The dynamical and radiative properties of the quiescent state (X-ray luminosity $\lesssim10^{34}\ \ergs$) of black hole X-ray transients (BHXTs) remains unclear, mainly because of low-luminosity and poor data quantity. We demonstrate that, the simultaneous multi-wavelength (including radio, optical, ultraviolet and X-ray bands) spectrum of V404 Cyg in its bright quiescent state can be well described by the radiation from the companion star and more importantly, the compact jet. Neither the outer thin disc nor the inner hot accretion flow is important in the total spectrum. Together with recent findings, i.e. the power-law X-ray spectrum and the non-variable X-ray spectral shape (or constant photon index) in contrast to the dramatic change in the X-ray luminosity, we argue the quiescent state spectrum of BHXTs is actually jet-dominated. Additional observational properties consistent with this jet model are also discussed as supporting evidences.
\end{abstract}

\begin{keywords}
accretion, accretion discs - black hole physics - X-rays: binaries - stars: jets - stars: individual (V404 Cyg)
\end{keywords}

\section{Introduction}\label{intro}
Soft X-ray transients are binary systems, where the primaries (black holes or neutron stars) accrete material from their companion low-mass stars through Roche lobe overflow. The black hole X-ray transients generally undergo occasional outbursts, during which they exhibit distinctive states (soft, hard and intermediate) according to their spectral and timing properties \citep{zg04,hb05,rm06,d07,b10,zh13}, likely the consequence of the changes in geometry and radiation mechanism of the accretion flow (e.g. \citealt{e97}). The soft state can be well described by cold disc model (\citealt{ss73}, hereafter SSD). The hard state is now generally understood under the hot accretion-jet scenario (see \citealt{yn14} [YN14 hereafter] for an up-to-date review on hot accretion flow and its applications on various objects including BHXTs), developed from the truncated disc model originally proposed by \citet{e97}. Three components are involved in this model, 1) an outer SSD, which is truncated at radius $R_{\rm tr}$, 2) an inner hot accretion flow within such radius, and 3) a relativistic jet. The hot accretion flow generally stands for the radiativelly inefficient, advection-dominated accretion flow (ADAF; \citealt{ny94}), and it is updated by the recent progresses in accretion theory (see \citealt{xy12}; YN14 for summaries), i.e. the existence of outflow and the direct viscous heating to electrons. This model provides comprehensive explanations on the spectral and timing properties of BHXTs in their hard states (e.g. \citealt{y05a} [YCN05 hereafter]).

For majority of the time BHXTs are extraordinarily faint, characterised by the so-called ``quiescent state'', with the X-ray ($2-10\ {\rm keV}$) luminosity $L_{\rm X} \lesssim 10^{34} \ergs$. Mainly because of the difficulties in detection, the nature of the quiescent states remains unclear \citep{n02,nm08}. For example, although the origin of the X-ray in quiescent state should come from the high-energy electrons near black holes, it is still unclear whether it is the synchrotron radiation from the non-thermal electrons in the jet (e.g. \citealt{f03,y05b}), or the synchrotron self-Comptonization from the thermal electrons in ADAF (e.g. \citealt{n97}), or Comptonization (in jet) of seed photons from cold disc \citep{m05,k08}. From the accretion-jet model, it is shown that \citep{y05b}, because the ADAF emission in X-rays decreases faster than the jet emission with decreasing accretion rate $\mdot$, the X-ray radiation $L_{\rm X}$ may be dominated by the jet emission when $\mdot$ is below a certain threshold. In other words, the X-ray emission of the quiescent state of BHXTs is likely to from the jet. This prediction has passed several observational tests, mainly through spectral fitting (\citealt{p08}; YCN05). However, these tests suffer the shortage of poor observational data (the data are non-simultaneous and/or limited to narrow wavebands), thus lead to some debates.

We in this {\it Letter} aim to argue that the quiescent state of BHXTs can be well-characterised by the jet model. We first in \S \ref{jetmodel} give a brief description of the jet model we used, and then in \S \ref{specfit} provide a comprehensive broad band (from radio to X-ray) spectral fitting of the quiescent spectrum of V404 Cyg, where simultaneous X-ray, ultraviolet (UV), optical and radio observations are available. We then in \S \ref{evidence} provide additional observational supports. Finally we provide some discussions and a brief summary in \S \ref{summary}.

\section{Jet model}\label{jetmodel}

Our jet model is phenomenological (see YCN05 for more details). The composition is assumed to be normal plasma (cf. \citealt{s11} for related discussions). The bulk velocity is assumed to be small, $V_{\rm jet}\approx 0.55\ c$, i.e. bulk Lorentz factor of the compact jet is $1.2$ (typical for the jets in the hard state of BHXTs; \citealt{f06}).

Internal shocks within the jet will accelerate a fraction $\xi$ of the electrons into energy distribution of power law, with the index of $p_{\rm jet}$. Due to the strong radiative cooling, the high-energy part of the accelerated power-law electrons will be cooled down, and their distribution index will be $p_{\rm jet}+1$ \citep{rl79}. Two additional parameters, $\epsilon_e$ and $\epsilon_B$, are also included to quantify the fraction of the shock energy that goes into electrons and magnetic fields, respectively. For simplicity, all these microphysical parameters are assumed to be constant along the jet direction.

With above parameters given, the synchrotron emission from these accelerated electrons can be calculated. Effects of inverse Compton scattering of the synchroton photons in the jet are negligible due to small scattering optical depth (but see \citealt{m05}). Generally speaking, the spectral energy distribution (SED) of the jet (or more generally the power-law electrons) is fairly simple. The high energy part (e.g. UV and X-ray bands) is power-law, and because the electrons responsible for the X-ray radiation are actually cooled, the photon index is $\Gamma \equiv 1-\alpha \approx 1+(p_{\rm jet}+1-1)/2 = 1+p_{\rm jet}/2$ (\citealt{rl79}. Spectral index $\alpha$ is defined through $F_\nu \propto \nu^\alpha$, where $F_\nu$ is the flux). The low energy part (e.g. radio up to IR) is also power-law, because of self-absorption, and the spectrum is flat or slightly inverted with spectral index $\alpha \approx 0 - 0.5$.

Generally degeneracies in parameters exist for the modelling. $\mdot_{\rm jet}$ (the mass loss rate into jet) is sensitively coupled with $V_{\rm jet}$ (fixed in all our applications) which controls beaming effect and gas density. Besides, the radiation at every waveband relates positively to parameters $\mdot_{\rm jet}$, $\epsilon_e$ and $\epsilon_B$. Moreover, parameter $\epsilon_e$ is more sensitive to the radiation at high-$\nu$ (e.g. X-ray and UV) energy band compared to that at low-$\nu$ (e.g. radio and IR), while $\epsilon_B$ (also $\mdot_{\rm jet}$, with weaker effects) shows opposite effects, i.e. more sensitive at low-$\nu$ bands. Parameter $\xi$, on the other hand, plays a more complex role. Increasing it (i.e, reducing the average energy of the power-law electrons in the jet) will reduce the X-ray radiation while enhance the radio emission. Finally, the X-ray spectral shape is mainly determined by $p_{\rm jet}$ (cf. Fig.\ \ref{jetzone}), and affected also by $\xi$ and $\epsilon_e$.


\section{SED of the quiescent state of V404 Cyg}\label{specfit}

\begin{figure}
\centerline{\includegraphics[width=8.5cm]{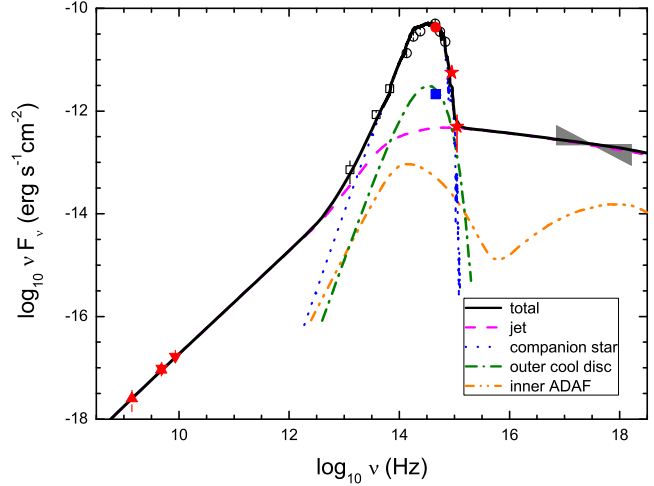}}
\caption{Quiescent SED of V404 Cyg. The solid square indicates the optical flickering emission. The simultaneous and non-simultaneous data are respectively shown as solid and open marks (data from \citealt{h09}). }
\label{V404_SED}
\vspace{-0.3cm}
\end{figure}

V404 Cyg, located at a distance of $2.39\pm0.14~{\rm kpc}$ \citep{mj09}, is a black hole binary with a 6.5-d orbital period. The black hole mass is $M_{\rm BH} = 12 \pm 2\ \msun$, and the inclination of the system is $\theta=56\ {\rm degrees}$ (\citealt{s94}). The companion star is a low-mass K0 {\sc iii}-type giant star with an effective temperature $T_{\rm eff} \simeq 4570\ {\rm K}$ \citep{c94,s94}. This BHXT has the highest quiescent X-ray luminosity to date, i.e. $L_{\rm X}\approx 1.6\times10^{-7} \ledd$ ($\ledd$ is the Eddington luminosity). Besides, because the BHXTs are known to be variable even within one day, simultaneous observations are crucial to reduce the uncertainties in modelling. V404 Cyg, which also have most broad simultaneous observations of the quiescent state, turns out to be the best candidate for the investigation of the properties of quiescent states.

As shown in Fig.\ \ref{V404_SED}, we use observations carried out on 2003 July 28-29, which provide strict simultaneous data (shown as solid marks) for radio (by VLA and WSRT), optical (by IAC80, WHT and Gemini), UV (by {\it Hubble}) and X-ray (by {\it Chandra}).  For a better SED coverage, we supplement the spectrum with non-simultaneous data (shown as open marks) in infrared (IR; by UKIRT) and optical (by WHT) bands, which are observed on 1990 June 10, and in IR band observed by {\it Spitzer} in 2004-2005 \citep{g07}. Interested readers are referred to \citet{h09} for details of data reduction and archival compilation. The quiescent state of this source was also modelled under the jet model by \citet{p08} with a focus on the high-quality {\it XMM-Newton} data, and under pure ADAF model by \citet{n97}.

We apply the jet model, together with the emission from the companion star, to fit the broad SED. Stellar atmosphere model \citep{k93} is applied for the radiation of the companion (see \citealt{h09} for details). Evidently the companion dominates the radiation in mid-IR-optical bands (cf. dotted curve in Fig.\ \ref{V404_SED}).

The X-ray photon index is observationally derived as $\Gamma=2.17_{-0.14}^{+0.12}$ \citep{h09}. We thus pre-fix in our jet model $p_{\rm jet}=2.4$, a reasonable value according to the diffusive shock acceleration theory. The UV emission, together with the X-ray flux, provide an upper limit on $p_{\rm jet}$ parameter. This is because, for a given X-ray luminosity, larger $p_{\rm jet}$ will lead to softer optically thin synchrotron radiation, thus higher flux at lower energy bands.

\textcolor[rgb]{1.00,0.00,0.00}{There are four free parameters left, $\mdot_{\rm jet}, ~\xi, ~\epsilon_e$ and $\epsilon_B$. Through spectral modelling (i.e. radio flux and its slope, X-ray and UV fluxes)}, we find that, $\mdot_{\rm jet} = 3 \times 10^{-6} \mdot_{\rm Edd}$ ($\mdot_{\rm Edd} \equiv 10 \ledd/c^2$ the Eddington accretion rate). Additionally, $\xi = 0.08$, $\epsilon_e = 0.04$ and $\epsilon_B = 0.03$, all are within the typical range obtained in GRB afterglow (YCN05). The result is shown in Fig.\ \ref{V404_SED}. We find that the radiation in mid-UV up to X-ray bands is the optically thin part of the synchrotron radiation. On the other hand, the radiation in radio up to mid-IR bands is the optically thick self-absorbed synchrotron radiation, and the spectrum is flat with spectral index $\alpha_{R-IR} \approx 0.02$. Most remarkably, as shown in Fig.\ \ref{V404_SED}, the UV flux (at $\nu\approx 1.1\times10^{15}\ {\rm Hz}$) simultaneously observed by {\it Hubble}, which is more than one order of magnitude lower compared to the prediction of a pure ADAF model that produces the same X-ray luminosity (see e.g. \citealt{h09}), is highly consistent with the jet model.

Now we turn to the turnover frequency, which characterises the location between optically thin and optically thick synchrotron radiation. This frequency is derived recently in several BHXTs \citep{g11,r13}, where they intend to constrain from this value the basic properties of jet, i.e. magnetic field strength and electron number density, etc. In quiescent state of V404 Cyg, we measured from our spectral fitting method a turnover frequency $\sim 1\times10^{14}{\rm Hz}$. There are two cautions we would like to address. {\it First,} the turnover frequency is sensitive to the properties of the inner central regions of the jet, where the local turnover frequency is the highest (cf. Fig.\ \ref{jetzone}). Correspondingly, the derived physical properties also reveal the properties of the central jet regions. {\it Second,} the application of using IR-UV bands for the extrapolation of the optically thin part should be cautious during the quiescent state, where the contamination from the companion star dominates. Even during the hard state whose the IR/optical emission is not from the companion, the contribution from the outer SSD in these bands still need to be well constrained/modelled and subtracted before any serious extrapolations of the near-IR-optical band. The far-IR could be better, but the turnover frequency is likely to be higher than far-IR. For the quiescent state, since we argue the X-ray is the optically thin synchrotron radiation, we recommend to use extrapolations between X-ray spectrum and radio spectrum to constrain the turnover frequency.

We now check the radiative contribution from the inner hot accretion flow and the outer cold disc. For this purpose, we assume the accretion rate at $5\ R_{\rm s}$ (here $R_{\rm s}$ is the Schwardzschild radius of the black hole) is $5\ \mdot_{\rm jet}$, i.e. $20\%$ of the gas near $5\ R_{\rm s}$ will enter into the jet. Basic parameters adopted (fixed, without any modelling) for the hot accretion flow are, outflow strength parameter $s=0.6$ ($\dot{M}\propto R^s$; see below for the constrain on this parameter), viscous parameter $\alpha_{\rm vis}=0.3$, gas to magnetic pressure ratio $\beta=9$, and the electron viscous heating fraction $\delta=0.1$. The numerical result for the inner ADAF is shown as double-dot dashed curve in Fig.\ \ref{V404_SED}, where the lower peak is the synchrotron and the second peak is bremsstrahlung (dominated) with a negligible fraction of inverse-Comptonization. Clearly from this plot the ADAF has no radiative importance at any wavebands for the quiescent state of BHXTs.

The remaining component is the outer irradiated cool disc. Generally its radiation is difficult to constrain in quiescent states, due to the dominance of the companion star. For V404 Cyg, \citet{h09} argued that the fast variable optical flickering (solid square in Fig.\ \ref{V404_SED}) may comes from the outer cool disc. In order to model this optical emission without exceeding the UV constraint, the truncated radius is found to be $R_{\rm tr} > 1.5\times10^4\ R_{\rm s}$\footnote{We note that a truncated radius similar to ours, $R_{\rm tr}\sim 10^4 R_{\rm s}$, is constrained in the quiescence of XTE J1118+480, whose UV extinction is low because of high latitude above the Galactic plane (cf. \citealt{m03}).}. Below we take $R_{\rm tr} = 2\times10^4\ R_{\rm s}$. Correspondingly, the accretion rate of the cool disc is $\ge 2.5\times 10^{-3} \mdot_{\rm Edd}$ (dot-dashed curve). We note that \citet{l00} provide a formulae to estimate the mass supply in the quiescent state as, $\mdot(R) \approx 4.0\times10^{15} \left(M_{\rm BH}/1\ \msun\right)^{-0.88}\ \left(R/10^{10}\ {\rm cm}\right)^{2.65}\ {\rm g\ s^{-1}}$. Applying this formulae to V404 Cyg, we find the accretion rate at $R_{\rm tr}$ is $\approx5\times10^{-3}\ \mdot_{\rm Edd}$, in good agreement with our constraint. With accretion rates at $5\ R_{\rm s}$ and $R_{\rm tr}$ provided, the outflow strength of the inner ADAF is constrained as $s \approx 0.6$, consistent with the suggested range ($s \sim 0.4-0.8$) from recent large-scale numerical simulations of hot accretion flows (\citealt{y12} and references therein).

We thus conclude, from detailed spectral fitting of the simultaneous SED, that emission from the compact jet and the companion dominates the whole spectrum during the quiescent state of BHXTs (at least for V404 Cyg). The predicted critical X-ray luminosity $L_{\rm X,crit}$ for the jet dominance of this source, depending on the unconstrained parameters of ADAF, can roughly be estimated to be $\sim 1.5-5\times10^{33}\ergs\sim 1-3\times10^{-6} \ledd$ (see also \citealt{y05b,y09,p13}), a factor of $\sim 6-20$ larger to the observed value. We note that additional examples of the jet-dominance can been found in low-luminosity active galactic nuclei \citep[AGNs; see e.g.][]{w07,p07,w08,y09,d11}.

\section{Additional evidences for the jet model}\label{evidence}

\subsection{strong evidences in favor of the jet-dominated model}

Besides the excellent spectral fitting of the simultaneous broad band SED in the quiescent state, there are several additional proofs on the jet-dominated quiescent state model, which we list below.

{\it First, the optically thick, flat or slightly inverted, radio spectrum.} The multi-band simultaneous radio observations of quiescent (and also hard) state show a flat spectrum \citep{g05}, with the spectral index $\alpha\sim 0$. This is usually taken as most obvious and {\it direct} evidence for the existence of compact relativistic jet, by the analogy to the jets in AGNs which are indeed imaged to show collimated conical structures. The radio (up to IR) radiation is dominated by the jet emission, where the flat or inverted spectrum is due to the combination of radiation from different locations (with different break frequency) along the jet direction (YCN05;\citealt{m05,k08}). Besides, the radio-IR emission is also expected to be polarized, which currently only observed during the hard states (see \citealt{f06}). The non-detection of radio polarization in the quiescent state \citep{mj08} may only because of weaker magnetic field strength compared to that in the hard state.

{\it Second, the power-law shape of the X-ray spectrum.} One notable difference in X-ray band between the ADAF-origin and the jet-origin is that, ADAF (at low accretion rate) predicts a curved bump-shaped spectrum (YN14, but see \citealt{pv14} and references therein for hybrid models considering the non-thermal electrons in ADAF), while the jet has a power-law spectrum (cf. Fig.\ \ref{V404_SED}). However, it is challenging to reliably determine the spectral shape of BHXTs in their quiescent states, mainly because of their faintness. \citet{p08} recently carried out deep high-quality {\it XMM-Newton} observations of three BHXTs, the exposure time for each source is more than $40\ {\rm ks}$. Two of them (GRO J1655-40 and V404 Cyg) are confirmed to have precisely power-law X-ray spectra, without any possible curvatures. The left one (XTE J1550-564), although not confirmed solidly due to relatively poorer data quality, is also most likely power-law.

{\it Third, constant photon index as X-ray luminosity varies by orders of magnitude.} The jet model {\it predicts} that, the X-ray spectral shape ($\Gamma$) is governed by the energy distribution of non-thermal electrons ($p_{\rm jet}+1$), and it will be insensitive to the X-ray luminosity, provided the acceleration mechanism leads to a convergent value on $p_{\rm jet}$. There are several investigations on the spectral changes in the quiescent states. \citet{p13} (see also \citealt{r14}) recently combined a sample of 10 BHXTs in their quiescence, with X-ray luminosities range in 4 orders of magnitude, and found that the $\Gamma$ plateaus to an average $\langle\Gamma\rangle\approx 2.1$. Besides this statistical work, \citet{p13} also carried out {\it Chandra} observations of three BHXTs in their decay phase, and found no evidence for the spectral change of each individuals (see also \citealt{c06}). Most remarkably, the photon index $\Gamma$ of MAXI J1659-152 is consistent with a constant value over 4-order-of-magnitude change of its X-ray luminosity. Besides, from a long-term (about 2 months) {\it Swift}/XRT monitoring of V404 Cyg during its quiescent state, \citet{bc14} find that its spectral shape remains roughly a constant as the flux varies (by a factor of $\sim 5$). The photon index is also constrained to be $\Gamma\approx2.10-2.35$, consistent with the prediction of the jet model. Similar conclusion is also derived by \citet{h09}, where they use the hardness (the flux ratio between hard and soft X-rays) as a diagnose of spectral properties, and find no variation in the hardness for a factor of $\sim10$ variation in X-ray luminosity.

This null $\Gamma-L_{\rm X}$ correlation is different from the anti-correlation usually observed in the hard states of BHXTs (e.g. \citealt{y07,wg08}), which was expected in ADAF model \citep{e97,e98,ql13}. Note that it remains unclear if the null correlation can also be understood by the maximal jet model \citep{m05,k08}. Furthermore, the X-ray spectrum of quiescent states ($\Gamma \approx 2.1-2.3$) is much softer compared to that of hard states (typically $\Gamma \approx 1.6$). These two observations indicate that the quiescent state could be different dynamically and/or radiatively from the hard state.

\begin{figure}
\centerline{\includegraphics[width=6.5cm]{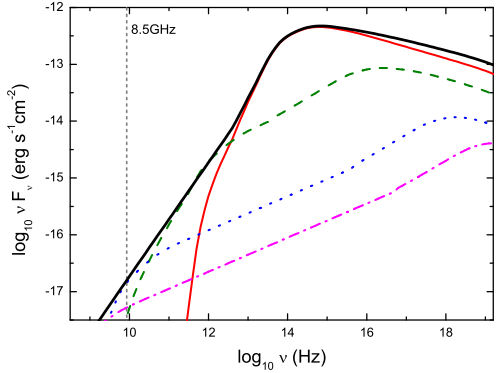}}
\caption{Emission from different height ($z$) along the jet direction. The thick solid line is the total jet radiation. The solid, dashed, dotted, and dot-dashed curves are, respectively, for height $z/R_{\rm s}$ in the range $10 - 10^3$, $10^3 - 10^5$, $10^5 - 10^6$, and $>10^6$.}
\label{jetzone}\vspace{-0.3cm}
\end{figure}

\subsection{observational features consistent with the jet model}
Apart from the solid evidences listed above, there are also observational properties that although can also be understood by other models, but are totally consistent with the jet model here. We below list some of them as supporting evidences.

{\it First, the size of the radio-emitting plasma.} Very few sources have sensitive radio images in their quiescent state. One notable exception is V404 Cyg, which was observed at $8.5\ {\rm GHz}$ on Dec. 2 2007 by VLA together with additional radio telescopes like Green Bank Telescope and Effelsberg. From the (unresolved) radio image, the size of the radio-emitting plasma is constrained to be less than $1.3\ {\rm mas}$ or equivalently $2\times10^{7}\ R_{\rm s}$ \citep{mj08}.

The flux of this observation, $(0.2-0.4)\ {\rm mJy}$ (non-flaring emission), is similar to the flux we modelled in \S \ref{specfit}, thus the upper limit may also apply to the observation we adopted. We shown in Fig.\ \ref{jetzone} (see also Fig. 5 in \citealt{z14}) the emission from different locations along the jet direction. We find that most ($>90\%$) of the radiation at $8.5\ {\rm GHz}$ comes from the location between $10^5\ R_{\rm s}$ and $10^6\ R_{\rm s}$, or a size of $\sim 0.07\ {\rm mas}$ at $2.4\ {\rm kpc}$. This is consistent with the observational upper limit. Our theoretical result also indicates that extremely-high resolution radio arrays, i.e. with space-VLBI, are required in order to resolve these compact jets in BHXTs, a huge challenge to the current/future radio facilities.

{\it Second, the variation in radio and X-ray bands.} In the quiescent state, radio emission is found to be much less variable compared to the X-ray emission (for V404 Cyg, see e.g. \citealt{c08,h09}). This feature could be naturally understood under the jet model. As illustrated in Fig.\ \ref{jetzone}, although both radio and X-ray wavebands are dominated by emission from jet, they actually originate from different locations along the jet. The X-ray emission comes from the inner (compact) regions of jet, while the radio emission mainly comes from the outer (broad) regions. Since the size of emission regions provides the lower limit for the variability timescale (fluctuations will be smoothed out within timescale $\sim z/c_s$, where $c_s$ is the sound speed of the electrons), the weaker variability in radio is totally expected.

Besides, it is also reported that there is no direct evidence of correlation for the variations between radio and X-ray \citep{h09}. This may also because of different radiative zones involved, and the variations are due to some local effect, e.g. local fluctuation in the occurrence efficiency of microphysics like magnetic reconnections and/or internal shocks, which has no direct causal connection to other zones.

{\it Third, the correlation in optical and X-ray variation.} In V404 Cyg, X-ray variability correlates well with H$\alpha$ \citep{h04} and optical continuum \citep{h09}, consistent with the irradiation/reprocessing model. H$\alpha$ also shows double-peaked profile \citep{h04}, indicates a cool-disc-origin. All these are usually seen in bright hard X-ray states, where the accretion-jet model being successfully applied. The jet-dominated quiescent model is a naturally extension of that model, the only dynamical/structural difference is that the outer disc retreats further away. Besides this structural difference, the UV-X-ray photons to irradiate the disc is also different. i.e. an ADAF origin in bright hard state and a jet origin in quiescent state. However, since ADAF and the jet are tightly correlated, and the location of X-ray photons is the central region in both cases, similar correlations should be expected.

Besides, the spectral fitting provides a constraint on $R_{\rm tr}$, $\ge 1.5\times10^4 R_{\rm s}$. Optical observations on different epochs (2002 Sep.) by ULTRACAM spotted a $0.78\ {\rm mHz}$ quasi-periodic-oscillation (QPO) feature. If the optical variation relates to the SSD (ref. Fig.\ \ref{V404_SED} and \S \ref{specfit}) and the QPO relates with the orbital frequency (Keplerian) at $R_{\rm tr}$ \citep{gs04}, then we have $R_{\rm tr}\sim 1.1\times10^{4}\ R_{\rm s}$ (\citealt{s03}, see also Narayan et al. 1997.).

\section{Discussions and Summary}\label{summary}


One consequence of the jet-dominated quiescent state is that, the radio/X-ray luminosity relationship steepens to $b\approx 1.23$ (where $L_{\rm R} \propto L_{\rm X}^b$; \citealt{y05b}), compared to $b\approx 0.62$ typically observed in the hard states of BHXTs \citep{c03,c13}. However, such steepen has not been observed in BHXTs yet. For V404 Cyg, our spectral modelling indicates it enters the steep correlation regime and the radio flux will be $(3-6)$ times lower to that extrapolated from the $b\sim 0.6$ correlation, a difference not be observed \citep{c08}. Further efforts are still needed to understand this discrepancy. On the other hand, we note that observationally the low-luminosity AGNs (which are analogy to BHXTs in their hard and quiescent states) do exhibit a steeper correlation, with $b\approx 1.22$, in excellent agreement with the theoretical prediction \citep{y09}.


Besides, if the X-ray radiation of the quiescent state has a jet-origin, then, we would expect to have relatively high degree of polarization in X-ray band, compared to the typical hard state, where the X-rays are of inverse-Compton origin. Currently the X-ray polarization is nearly an unexplored field in astronomy, and future sensitive spaceborne X-ray polarimeters (e.g. the proposed X-ray Timing and Polarization [XTP] mission and the Gravity and Extreme Magnetism SMEX X-ray polarimetry mission) may detect variable X-ray polarization from synchrotron emission of the jets.


We now give a brief summary of this work. With recent advances in X-ray observations, especially the deep sensitive spectral observation and the long-term monitoring, we confirm that the emission of the quiescent state of BHXTs is dominated by the radiation from the compact relativistic jet. The outer thin disc and the inner hot accretion flow generally play negligible roles in radiation, while the companion dominates the emission between mid-IR and optical bands. We also illustrate (cf. \S \ref{evidence}) that the jet-dominated quiescent state model can explain most of the observational features.

\vspace{-0.5cm}
\section*{Acknowledgments}
We thank Xuebing Wang (SHAO) for the stellar emission calculation. This work was supported in part by the Natural Science Foundation of China (grants 11133005, 11121062, 11203057 and 11333004), the 973 Program (grant 2014CB845800), and the Strategic Priority Research Program ``The Emergence of Cosmological Structures'' of CAS (grant XDB09000000).

\label{lastpage}


\vspace{-0.5cm}
\begin{thebibliography}{}

\bibitem[\protect\citeauthoryear{Belloni}{2010}]{b10} Belloni T.~M., 2010, in ``The Jet Paradigm - From Microquasars to Quasars'', ed. T. Belloni,  Lecture Notes in Physics, Springer-Verlag, Berlin, 794, 53

\bibitem[\protect\citeauthoryear{Bernardini \& Cackett}{2014}]{bc14} Bernardini F., Cackett E. M., 2014, \mnras, 439, 2771


\bibitem[\protect\citeauthoryear{Casares \& Charles}{1994}]{c94} Casares J., Charles P. A., 1994, \mnras, 271, L5

\bibitem[\protect\citeauthoryear{Corbel et al.}{2013}]{c13} Corbel S., Coriat M., Brocksopp C., Tzioumis A.~K., et al., 2013, \mnras, 428, 2500

\bibitem[\protect\citeauthoryear{Corbel et al.}{2003}]{c03} Corbel S., Nowak M.~A., Fender R.~P., Tzioumis A.~K., Markoff S., 2003, A\&A, 400, 1007

\bibitem[\protect\citeauthoryear{Corbel, Tomsick \& Kaaret}{2006}]{c06} Corbel S., Tomsick J.~A., Kaaret P., 2006, \apj, 636, 971

\bibitem[\protect\citeauthoryear{Corbel et al.}{2008}]{c08} Corbel S., Koerding E., Kaaret P., 2008, \mnras, 389, 1697

\bibitem[\protect\citeauthoryear{de Gasperin et al.}{2011}]{d11} de Gasperin F., Merloni A., Sell P., et al., 2011, \mnras, 415, 2910

\bibitem[\protect\citeauthoryear{Done, Gierlinski \& Kubota }{2007}]{d07} Done, C., Gierlinski, M., Kubota, A., 2007, A\&ARv, 15, 1

\bibitem[\protect\citeauthoryear{Esin et al.}{1997}]{e97} Esin A.~A., McClintock J.~E., Narayan R., 1997, \apj, 489, 865

\bibitem[\protect\citeauthoryear{Esin et al.}{1998}]{e98} Esin A.~A., Narayan R., Cui W., Grove J.~E., Zhang S.~N., 1998, \apj, 505, 854

\bibitem[\protect\citeauthoryear{Fender}{2006}]{f06} Fender R.~P., in ``Compact stellar X-ray sources'', Eds. W. Lewin \& M. van der Klis, Cambridge Astrophysics Series, No. 39. Cambridge, UK: Cambridge University Press, 2006, 381

\bibitem[\protect\citeauthoryear{Fender, Gallo \& Jonker}{2003}]{f03} Fender R.~P., Gallo E., Jonker P.~G., 2003, \mnras, 343, L99

\bibitem[\protect\citeauthoryear{Gallo, Fender \& Hynes}{2005}]{g05} Gallo E., Fender R.~P., Hynes R.~I., 2005, \mnras, 356, 1017

\bibitem[\protect\citeauthoryear{Gallo et al.}{ 2007}]{g07} Gallo E., Migliari S., Markoff S., Tomsick J.~A. et al., 2007, \apj, 670, 600

\bibitem[\protect\citeauthoryear{Giannios \& Spruit}{2004}]{gs04} Giannios, D., Spruit, H.~C. 2004, A\&A, 427, 251

\bibitem[\protect\citeauthoryear{Gandhi et al.}{2011}]{g11} Gandhi P., Blain A.~W., Russell D.~M., Casella P., et al., 2011, \apj, 740, L13

\bibitem[\protect\citeauthoryear{Homan \& Belloni}{2005}]{hb05} Homan, J., Belloni, T., 2005, Ap\&SS, 300, 107


\bibitem[\protect\citeauthoryear{Hynes et al.}{2004}]{h04} Hynes R.~I., Charles P.~A., Garcia M.~R., Robinson E.~L. et al., 2004, \apj, 611, L125

\bibitem[\protect\citeauthoryear{Hynes et al.}{2009}]{h09} Hynes R.~I., Bradley C.~K., Rupen M., Gallo E. et al., 2009, \mnras, 399, 2239

\bibitem[\protect\citeauthoryear{Kurucz}{1993}]{k93} Kurucz R.~L., 1993, SYNTHE Spectrum Synthesis Programs and Line Data (CD-ROM), Smithsonian Astrophysical Observatory, Cambridge, MA

\bibitem[\protect\citeauthoryear{Kylafis et al.}{2008}]{k08} Kylafis N.~D., Papadakis I.~E., Reig P., Giannios D., Pooley G.~G., 2008, A\&A, 489, 481

\bibitem[\protect\citeauthoryear{Lasota}{2000}]{l00} Lasota J.~P., 2000, A\&A, 360, 575

\bibitem[\protect\citeauthoryear{Markoff, Nowak \& Wilms}{2005}]{m05} Markoff S., Nowak M.~A., Wilms J., 2005, \apj, 635, 1203

\bibitem[\protect\citeauthoryear{McClintock et al.}{2003}]{m03} McClintock J.~E., Narayan R., Garcia M.~R., Orosz J.~A., et al., 2003, \apj, 593, 435

\bibitem[\protect\citeauthoryear{Miller-Jones et al.}{2008}]{mj08} Miller-Jones J.~C.~A., Gallo E., Rupen M.~P., Mioduszewski A.~J., et al., 2008, \mnras, 388, 1751

\bibitem[\protect\citeauthoryear{Miller-Jones et al.}{2009}]{mj09} Miller-Jones J.~C.~A., Jonker P.~G., Dhawan V., Brisken W., et al. 2009, ApJ, 706, L230

\bibitem[\protect\citeauthoryear{Narayan, Garcia \& McClintock}{2002}]{n02} Narayan R., Garcia M.~R.,  McClintock J.~E., 2002, in The Ninth Marcel Grossmann Meeting, eds. V.~G. Gurzadyan, R.~T. Jantzen, R. Ruffini (Singapore: World Scientific), 405

\bibitem[\protect\citeauthoryear{Narayan \& McClintock}{2008}]{nm08} Narayan R., McClintock J.~E., 2008, NewAR, 51, 733

\bibitem[\protect\citeauthoryear{Narayan, Barret \& McClintock}{1997}]{n97} Narayan R., Barret D., McClintock J.~E., 1997, \apj, 482, 448

\bibitem[\protect\citeauthoryear{Narayan \& Yi}{1994}]{ny94} Narayan R., Yi I., 1994, \apj, 428, L13

\bibitem[\protect\citeauthoryear{Pellegrini et al.}{2007}]{p07} Pellegrini S., Siemiginowska A., Fabbiano G., et al., 2007, \apj, 667, 749

\bibitem[\protect\citeauthoryear{Plotkin, Gallo \& Jonker}{2013}]{p13} Plotkin R.~M., Gallo E., Jonker P.~G., 2013, \apj, 773, 59

\bibitem[\protect\citeauthoryear{Poutanen \& Veledina}{2014}]{pv14} Poutanen J., Veledina A., 2014, Space Sci Rev. (in press)

\bibitem[\protect\citeauthoryear{Pszota et al.}{2008}]{p08} Pszota G., Zhang H., Yuan F., Cui W., 2008, \mnras, 389, 423

\bibitem[\protect\citeauthoryear{Qiao \& Liu}{2013}]{ql13} Qiao E. L., Liu B. F., 2013, \apj, 764, 2

\bibitem[\protect\citeauthoryear{Remillard \& McClintock}{2006}]{rm06} Remillard R.~A., McClintock J.~E., 2006, \araa, 44, 49

\bibitem[\protect\citeauthoryear{Reynolds et al.}{2014}]{r14} Reynolds M.~T., Reis R.~C., Miller J.~M., Cackett E.~M., Degenaar N., \mnras, in press (arXiv: 1405.0474)

\bibitem[\protect\citeauthoryear{Russell et al.}{2013}]{r13} Russell D.~M., Russell T.~D., Miller-Jones J.~C.~A., O'Brien K., et al., 2013, \apj, 768, L35

\bibitem[\protect\citeauthoryear{Rybicki \& Lightman}{1979}]{rl79} Rybicki G.~B., Lightman A.~P., 1979, Radiative Processes in Astrophysics (New York: Wiley)

\bibitem[\protect\citeauthoryear{Shahbaz et al.}{1994}]{s94} Shahbaz T., Ringwald F. A., Bunn J. C., et al. 1994, \mnras, 271, L10

\bibitem[\protect\citeauthoryear{Shahbaz et al.}{2003}]{s03} Shahbaz T., Dhillon V.~S., Marsh T.~R., Zurita C., et al. 2003, \mnras, 346, 1116

\bibitem[\protect\citeauthoryear{Shakura \& Sunyaev}{1973}]{ss73} Shakura N.~I., Sunyaev R.~A., 1973, A\&A, 24, 337

\bibitem[\protect\citeauthoryear{Sikora}{2011}]{s11} Sikora M., 2011, in IAUS No. 275, ``Jets at all Scales'', eds. G.E. Romero, R.A. Sunyaev, and T. Belloni, p. 59

\bibitem[\protect\citeauthoryear{Wrobel, Terashima \& Ho}{2008}]{w08} Wrobel J.~M., Terashima Y., Ho L.~C., 2008, \apj, 675, 1041

\bibitem[\protect\citeauthoryear{Wu \& Gu}{2008}]{wg08} Wu Q., Gu M., 2008, \apj, 682, 212

\bibitem[\protect\citeauthoryear{Wu, Yuan \& Cao}{2007}]{w07} Wu Q., Yuan F., Cao X., 2007, \apj, 669, 96

\bibitem[\protect\citeauthoryear{Xie \& Yuan}{2012}]{xy12} Xie, F. G., \& Yuan, F. 2012, \mnras, 427, 1580

\bibitem[\protect\citeauthoryear{Yuan, Cui \& Narayan}{2005}]{y05a} Yuan F., Cui W., Narayan R., 2005a, \apj, 620, 905

\bibitem[\protect\citeauthoryear{Yuan \& Cui}{2005}]{y05b} Yuan F., Cui W., 2005b, \apj, 629, 408

\bibitem[\protect\citeauthoryear{Yuan et al.}{2007}]{y07} Yuan F., Taam R. E., Misra R. et al., 2007, \apj, 658, 282

\bibitem[\protect\citeauthoryear{Yuan \& Narayan}{2014}]{yn14} Yuan F., Narayan R., 2014, \araa (in press), arXiv: 1401.0586

\bibitem[\protect\citeauthoryear{Yuan, Wu \& Bu}{2012}]{y12} Yuan F., Wu M., Bu D., 2012, \apj, 761, 129

\bibitem[\protect\citeauthoryear{Yuan, Yu \& Ho}{2009}]{y09} Yuan F., Yu Z., Ho L., 2009. \apj, 703, 1034

\bibitem[\protect\citeauthoryear{Zdziarski \& Gierlinski}{2004}]{zg04} Zdziarski A.~A., Gierlinski M., 2004, Prog. Theo. Phy. Supp., 155, 99

\bibitem[\protect\citeauthoryear{Zdziarski et al.}{2014}]{z14} Zdziarski A.~A., Pjanka P., Sikora M., Stawarz L., 2014, \mnras, arXiv: 1403.4768

\bibitem[\protect\citeauthoryear{Zhang}{2013}]{zh13} Zhang S.~N., 2013, Frontiers of Phy, 8, 630

\end{thebibliography}
\end{document}